# Beam broadening of polar molecules and clusters in deflection experiments

J. Bulthuis[1] and V. V. Kresin[2]

[1]Institute for Lasers, Life and Biophotonics, Vrije Universiteit, De Boelelaan 1083, 1081 HV Amsterdam, The Netherlands

[2]Department of Physics and Astronomy, University of Southern California, Los Angeles, California, 90089-0484, USA

*December 13, 2011*

## Abstract

A beam of rotating dipolar particles (molecules or clusters) will broaden when passed through an electric or magnetic field gradient region. This broadening, which is a common experimental observable, can be expressed in terms of the variance of the distribution of the resulting polarization orientation (the direction cosine). Here the broadening for symmetric-top and linear rotors is discussed. These two types of rotors have qualitatively different low-field orientation distribution functions, but behave similarly in a strong field. While analytical expressions for the polarization variance can be derived from first-order perturbation theory, for experimental guidance it is important to identify the applicability and limitations of these expressions, and the general dependence of the broadening on the experimental parameters. For this purpose, the analytical results are compared with the full diagonalization of the rotational Stark-effect matrices. Conveniently for experimental estimations, it is found that for symmetric tops the dependence of the broadening parameter on the rotational constant, the axial ratio, and the field strength remains similar to the analytical expression even outside of the perturbative regime. Also, it is observed that the shape envelope, the centroid, and the width of the orientation distribution function for a symmetric top are quite insensitive to the value of its rotational constant (except at low rotational temperatures).

# 1. Introduction

The last decade has seen a rejuvenation of interest in the study of polar molecules and nanoclusters in beams by electric and magnetic deflection (see, e.g., the reviews 1-3), a method first introduced and explored in the 1920s and 1930s.[4,5] The interest has been strengthened by the appearance of new techniques, such as brute force orientation,[6,7] laser induced alignment,[8] electric field deceleration and guiding,[9,10] state selection,[11,12] as well as interferometry and diffraction.[11,13] In principle, state selection by multipole focusing[14,15] also can be viewed as a deflection technique.

Here we address the passage of a beam of particles with a permanent dipole moment through an electric or magnetic field deflector. Two common, basic, experimental observables in measurements of this type are the average deflection and the broadening of the beam. These parameters are often used to deduce the magnitude of the particles' electric or magnetic moments. However, care is required because in such a deflection experiment the particle ensemble does not achieve thermal equilibrium within the external field, and consequently the textbook Langevin-Debye susceptibility[16] is not necessarily applicable. For example, a beam of rigid polar rotators with a body-fixed dipole may be assigned a rotational temperature distribution acquired prior to its entry into the field region (e.g., in the beam source), while its passage through the field is collision-free and sufficiently slow to be treated as adiabatic. We will refer to this situation as "adiabatic entry," to be distinguished from the textbook case of "in-field thermal equilibrium." In the end, the experimentalist's goal is to relate the magnitudes of beam deflection and broadening to the particles' inherent dipole moments and shapes.

For example, in a recent paper[17] Heiles $et\ al.$ employed such an approach for a detailed case study of $Ge_9$, $Ge_{10}$, and $Ge_{15}$ clusters. They discussed the relationship between deflection and broadening on the one hand (employing analytical approximations as well as classical and quantum simulations of rotational motion) and dielectric and structural properties on the other hand. Guided by quantum chemical predictions for the geometric shapes of specific clusters, they considered symmetric, asymmetric, and isomerically mixed populations.

In cases when reliable structural information is unavailable or uncertain, one has to resort to assumptions and approximations, and it becomes important to identify their applicability and limitations. How reliable are the analytical and perturbative treatments, and how strongly do inferences from deflection data depend on the various parameters: the rotational temperature, the symmetry, the dipole moment, and the strength of the deflecting field? Analogously to our earlier examination of average particle orientation,[18] here we consider these questions as they relate to beam broadening. In this aspect, the paper can be viewed as complementary to the aforementioned work by Heiles $et\ al.$[17]

To set the stage, recall that the average deflection of the beam, associated with a permanent dipole moment, is given by[3,19]

$$d_z = \frac{a(\partial \mathcal{E}/\partial z)}{mv^2} \mu_0 \bar{P}_1. \qquad (1)$$

The constant $a$ incorporates the apparatus geometric factors, $m$ and $v$ are the particle mass and velocity, and $\mu_0$ is the permanent dipole moment. It is assumed (as is typically the case in practice) that the field $\vec{\mathcal{E}} = \mathcal{E}\hat{z}$ and its gradient $\partial \mathcal{E}/\partial z$ are collinear. For a particular molecular quantum state $n$, $\bar{P}_1$ is the expectation value of the direction cosine of the dipole axis relative to the field:

$$P_1|_n \equiv \langle \cos\theta \rangle_n = \langle \mu_z \rangle_n / \mu_0. \tag{2}$$

The bar in Eq. (1) denotes averaging over the initial (prior to field entry) distribution of states in the beam ensemble:

$$\overline{P_1} = \overline{\langle \cos\theta \rangle_n} = \overline{\langle \mu_z \rangle_n} / \mu_0. \tag{3}$$

As stated above, consideration here will be limited to rigid systems (deflection of non-rigid clusters involves different considerations, see, e.g., Refs. 1-3,20,21), in which case the relevant quantum states $|n\rangle$ are the rotational ones $|JKM\rangle$, where $K$ and $M$ are quantum numbers corresponding to the projections of the rotational angular momentum $\vec{J}$ on the molecular z-axis and the laboratory Z-axis, respectively. Their distribution will be assumed to have the statistical Boltzmann form $\exp(-E_n^{rot}/k_B T_{rot})$. Note that the exponent contains only the rotational energies, as appropriate for the adiabatic-entry case: in the case of in-field equilibrium one would also include the Stark shift.

For concreteness, the notation and sample calculations in this paper refer to electric fields and electric dipole moments, but the results are equally applicable to the deflection of magnetic species in a Stern-Gerlach-type experiment. (Note, however, that due to inversion symmetry a spherical rigid rotor cannot have a finite electric dipole moment but may have a magnetic moment, since the former is a polar vector and the latter is an axial vector.)

It has been shown in earlier work[22,18] that the orientation for adiabatic entry is quite distinct from the well-known Langevin-Debye expression for in-field equilibrium. The latter corresponds to the universal form $\overline{P_1}^{L-D} = \frac{1}{3}\mu_0 \mathcal{E}/(k_B T_{rot})$ (in the weak-field limit), whereas for adiabatic entry second-order perturbation theory yields the following result for a polar symmetric top:

$$\overline{P_1} = \frac{\mu_0 \mathcal{E}}{k_B T_{rot}} z(\kappa). \tag{4}$$

The shape parameter $\kappa$ depends on the ratio of rotational constants:[23] $\kappa=(C-A)/C$ for a prolate top (with $B=C<A$ and therefore $\kappa<0$), and $\kappa=(A-C)/A$ for an oblate top (with $B=A>C$ and $0<\kappa<0.5$); $z(\kappa)$ is a function written out in Ref. [18]. For small deviations from a spherical rotor, i.e., for small $\kappa$, one finds $z(\kappa) = \frac{2}{9}\left(1 - \frac{1}{5}\kappa + ...\right)$; thus for a spherical rotor[24,25] $z(0)=2/9$.

As mentioned above, in the present paper we focus on another experimentally accessible parameter: the broadening of the beam by the external deflection field. The focus is not on instrumental broadening caused by a nonuniformity of the deflecting force, but on the inherent effect caused by the different orientations of the permanent dipoles in the beam ensemble. Thus, while beam deflection is determined by the *average* of the distribution of dipole moment projections on the field axis, beam broadening is determined by the *variance* of this distribution.

## 2. The variance of the orientation distribution

Inherent broadening of the beam by a deflecting field comes from the fact that individual particles in the beam have, on average, slightly different orientations relative to the field and therefore experience different deflecting forces. To parametrize this broadening, $b$, one can employ the variance of the dipoles' orientation in the beam ensemble that enters the field region:

$$b^2 = \left(\frac{a(\partial \mathcal{E}/\partial z)}{mv^2}\mu_0\right)^2 \mathrm{var}(P_1),\qquad(5)$$

where

$$\mathrm{var}(P_1) = \left[\overline{\left(P_1|_{JKM}\right)^2} - \left(\overline{P_1|_{JKM}}\right)^2\right].\qquad(6)$$

In an actual experiment with a beam of finite spatial width, the measured beam profile is a convolution of this broadening with the geometric cross-section of the beam. Since the variance of a convolution is the sum of two individual variances,[26] the practical implication is that if $\sigma_0$ is the variance of the undeflected beam profile, and $\sigma_b$ is the variance of the deflected profile, then

$$b^2 = \sigma_b^2 - \sigma_0^2.\qquad(7)$$

This links the experimentally measured beam broadening parameters with the underlying orientational variance of the individual dipoles.[3,27]

For a numerical evaluation, the energies and eigenfunctions of a dipolar rotor in the presence of a field can be calculated to high accuracy by diagonalizing the truncated Hamiltonian matrix.[18,28] For symmetric top molecules the electric field mixes only $J$ states, hence for each $K$ and $M$ (the states are doubly degenerate in $M$) one diagonalizes a matrix that includes all parent $J$ states populated at the given rotational temperature and a sufficient number of additional $J$ states to ensure convergence. (For asymmetric tops, which are not considered here, the calculation would be complicated by the fact that not only $J$, but also $K$ no longer remains a good quantum number.) Some results of such calculations, which in the following will be referred to as "exact," will be illustrated below, but for insight it is productive to begin with an analytic treatment based on perturbation theory.

Whereas the calculation of average orientation[18] required use of second-order perturbation theory (as mentioned in the Introduction), a finite expression for the variance can be obtained already in a first-order calculation. The general perturbation expansion for the orientation and the variance is written in terms of the dipole-field interaction parameter

$$\omega \equiv \mu_0 \mathcal{E}/B\qquad(8)$$

using the standard notation of Friedrich and Herschbach.[29] (Note that in Ref. 17 "$\omega$" is used to denote $\mu_0\mathcal{E}/k_B T_{\mathrm{rot}}$.) Numerically, for an electric dipole $\omega = 0.017\mu_0[\mathrm{D}]\mathcal{E}[\mathrm{kV/cm}]/B[\mathrm{cm}^{-1}]$, and for a magnetic dipole $\omega = 0.047\mu_0[\mu_B]\mathcal{H}[\mathrm{kG}]/B[\mathrm{cm}^{-1}]$.

Furthermore, since there is no finite orientation to first order, we can set $\overline{P_1|_{JKM}} = 0$ in Eq. (6) and therefore the expression for the broadening parameter reduces to

$$\mathrm{var}(P_1) \approx \overline{\left(P_1|_{JKM}\right)^2}.\qquad(9)$$

In contrast to the situation with beam deflection, differences in the broadening parameter between adiabatic-entry and in-field-equilibrium scenarios appear only if quadratic and higher-order terms in the field are included, hence in the low-field limit the results should be identical.

## 2.1 Symmetric top molecules and clusters

Using the first order perturbation expression

$$P_1|_{JKM} = \frac{MK}{J(J+1)}, \tag{10}$$

the variance of the dipole distribution is

$$\text{var}(P_1) \approx \frac{1}{Z_{rot}} \sum_{J=0}^{\infty} \sum_{K=-J}^{J} \sum_{M=-J}^{J} \frac{M^2 K^2}{J^2(J+1)^2} e^{-Y^{-1}\left[J(J+1) - \kappa K^2\right]}. \tag{11}$$

The shape parameter $\kappa$ was defined immediately following Eq. (4), and[29]

$$Y \equiv k_B T_{rot} / B. \tag{12}$$

The summation over $M$ can be factored out:

$$\text{var}(P_1) \approx \frac{1}{3Z_{rot}} \sum_{J=0}^{\infty} \sum_{K=-J}^{J} \frac{(2J+1)K^2}{J(J+1)} e^{-Y^{-1}\left[J(J+1) - \kappa K^2\right]}. \tag{13}$$

The rotational partition function is

$$Z_{rot} = \sum_{J=0}^{\infty} \sum_{K=-J}^{J} (2J+1) e^{-Y^{-1}\left[J(J+1) - \kappa K^2\right]}. \tag{14}$$

At this point, let us assume that a sufficient number of high-$J$ states are occupied to convert to integration over $K$ and $J$. To this end we replace $K$ by $[J(J+1)]^{1/2}\cos\varphi$, where $\varphi$ is the angle between $J$ and the dipole axis, and integrate over $0 \leq \varphi \leq 2\pi$ and then over $J$. This is the same procedure as used in Ref. 18. Conveniently, Eq. (13) turns out to have the same form as a term encountered in the evaluation of in-field average polarization [Eq. (21) in Ref. 18]. Thus one obtains for an oblate symmetric top ($0<\kappa<0.5$)

$$\text{var}(P_1) \approx \frac{1}{3\kappa}\left(1 - \frac{\sqrt{1-\kappa}}{\sqrt{\kappa}} \arcsin\sqrt{\kappa}\right) \tag{15}$$

and for a prolate symmetric top ($\kappa<0$)

$$\text{var}(P_1) \approx \frac{1}{3|\kappa|}\left(-1 + \frac{\sqrt{1+|\kappa|}}{\sqrt{|\kappa|}} \text{arcsinh}\sqrt{|\kappa|}\right). \tag{16}$$

Heiles *et al.*[17] derived the same equations in a slightly different way, by treating $J$, $K$ and $M$ as continuous variables.[30] (Note that they defined $\kappa$ in terms of the moments of inertia, instead of the rotational constants, so that the sign of this parameter in their work is reversed relative to our definition which follows earlier publications.[18,28])

Note that expressions (15) and (16) are independent of the temperature, the dipole moment, and the field strength, and depend only on the asymmetry parameter $\kappa$. This is a consequence of the fact that the leading order of $P_1|_{JKM}$, Eq. (10), does not contain the field interaction parameter $\omega$: the field dependence only enters in second order perturbation theory.

For $\kappa=0$, i.e., for a spherical top, the variance is

$$\text{var}(P_1) = \frac{1}{9}, \qquad (17)$$

in agreement with earlier results, see, e.g. Ref. 27.

Numerical examples in Fig. 1, based on exact quantum mechanical calculations for a symmetric top, confirm that for fields that are not too strong, var($P_1$) is determined only by the shape and agrees quantitatively with the analytical equations (15)-(17).

These equations, strictly speaking, should be applicable for small values of $\omega$, $Y^{-1}$ and $\mu_0\mathcal{E}/k_B T_{rot}=\omega/Y$. The plots in Fig. 1 show that for rising field strengths the variance begins to grow, and then decreases at very strong fields. This occurs as the orientation of different states initially diverges in an increasing field, and then converges with the onset of pendular state formation.

Additionally, as the rotational temperature increases, so does the number of $J$-states that contribute to the variance. Consequently, since higher $J$-states become pendular at a higher field strength, the peak of the var($P_1$) plot shifts to a higher field strength.

The general trends described here are independent of the ratio $A/C$. Even though it cannot have a permanent electric dipole moment, we included the case of a spherical top in the plots, for three reasons. First of all, it is a convenient conceptual reference point. Secondly, the calculation is identically applicable to magnetic dipoles, which can occur in spherical tops (see above). Thirdly, the actual asphericity of polar symmetric top clusters is often rather small (see e.g. Refs. 17,27).

It was mentioned above that the first-order variance should be identical for adiabatic field entry and for in-field equilibrium ensembles. The crosses in Fig. 1 confirm that the two ensembles yield close values up to very high field intensities.

Interestingly, Fig. 1 illustrates that the analytical expressions remain accurate even for values of the parameter $\omega$ (defined in Eq. 8) far beyond the $\omega \ll 1$ limit. For example, at a field strength of 10 kV/cm, $\omega=17$ for $B=0.01$ cm$^{-1}$ and $\mu_0=1$ D. We therefore conclude that $\omega$ is not a good measure for the applicability of the analytic expressions (15) and (16), in spite of the fact that second- and higher order perturbation terms contain powers of $\omega$. The second-order contribution to the variance, for instance, is proportional to $\omega^2$. Addition of this contribution does not yield to simple analytic evaluation, but it can be straightforwardly calculated numerically. As expected, and illustrated in Fig. 2(a), for $\omega<1$ (in the illustrated case $\omega=0.85$), this term makes a minor contribution to the variance, and brings it slightly closer to the exact value than the analytic first-order result. For $\omega>1$ (in the illustrated case $\omega=8.5$), however, the deviation from the exact calculation is strong. In fact, for $\omega>1$ second- and higher-order perturbation theory completely breaks down.

This is in stark contrast to the results of similar calculations of the orientation (Ref. 18), where the effects of applying higher than first-order perturbation terms are much smaller for comparable values of $\omega$.

The reason becomes clear if the effect of the second-order perturbation terms is considered in detail for pairs of interacting $J$ and $J+1$ levels. If the repulsion between interacting high- and low-field-seeking levels is overestimated, which is the case on average when applying second-order perturbations for $\omega \gtrsim 1$, the effect on the average orientation is limited: positive and negative orientations with the field tend to cancel. On the other hand, the effect on the variance may become large, and therefore for $\omega \gtrsim 1$ the addition of second-order terms is no alternative to an exact calculation of the variance.

The first-order perturbation, being diagonal, is independent of $\omega$ and does not give diverging results like the higher-order terms. While it falls short for increasing $\omega$ values, up to moderate values of this parameter the exact calculation does not deviate much from the first-order analytic result.

Fig. 2(b) and (c) illustrate the behavior of the variance as a function of the axial ratio according to the exact calculation, for varying rotational constants $B$ and electric field strengths $\mathcal{E}$. (The analytic first-order result is included as a reference.) Note that the curves retain a qualitatively similar form. This convenient fact makes it possible for one to estimate, in a deflection experiment, how strongly the beam broadening profile may be affected by particle shape variations.

Another convenient fact is that var($P_1$), like $\overline{P}_1$ itself,[18] is not especially sensitive to the magnitude of the rotational constant (as long as $k_B T_{rot}$ remains greater than $B$ to ensure that a sufficient number of rotational states are contributing to the deflection pattern). If quantum-mechanical rotational Stark effect matrices become prohibitively large for computation, it may be possible to artificially increase the value of $B$, thus decreasing the matrix size, without suffering a great loss of accuracy in the calculated deflection parameters. This observation, discussed further in Sec. 3.1, offers an alternative to performing a classical calculation.

The above is illustrated in Fig. 2(b), where the curves for $B$=0.002 cm$^{-1}$ and 0.1 cm$^{-1}$ nearly coincide. (The curve for $B$=0.01 cm$^{-1}$ has also been calculated but it cannot be distinguished from the $B$=0.002 cm$^{-1}$ curve.) Indeed, all these values fulfill the condition $B \ll k_B T$ for the assumed $T_{rot}$=5 K. On the other hand, $B$= 1 cm$^{-1}$ does not quite fulfill this condition (here $k_B T_{rot} \approx 3.5B$, so that only a few rotational states determine the variance) and its curve in Fig. 2(b) is indeed found to deviate from the others.

In Fig. 2(b) it is noticeable that the lowest-$B$ curves converge and deviate stronger from the first-order analytic (semiclassical) result than does the curve for $B$=1 cm$^{-1}$. Because the semiclassical result is based on the assumption that multiple rotational states contribute to the ensemble averages (so as to allow replacing summation by integration over states), one might initially expect the opposite, i.e., that it would be the curves for smaller $B$ that would lie closer to the exact result. The reason this is not the case is that for the small $B$ values the field interaction parameter has very large values (such as $\omega$=17 and $\omega$=850 for $B$=0.1 and 0.002 cm$^{-1}$, respectively).

Nevertheless, the conclusion above that $\omega$ by itself is not a key parameter for characterizing the variance remains valid in this case as well. Indeed, for increasing $\omega$ the interaction between rotational states gets stronger, but if at the same time $B$ decreases, then more high-lying rotational states contribute to the distributions. This turns out largely to counteract the effect of increasing $\omega$, so that all the curves retain a common shape, as pointed out above. Fig. 2(c) shows the dependence of the variance on the field strength $\mathcal{E}$ (i.e., effectively on $\omega$, since $B$ is kept constant in this case). If the field were increased still further, the variance (i.e., the beam broadening) would start to decrease, in accordance with Fig. 1.

Fig. 2 might be construed to suggest that the first-order analytic approximation always gives the lower limit of the variance. This is indeed the case as long as $k_B T$ does not decrease all the way to approach *B*. In the opposite limit, the variance will become dominated by the ground rotational state, and can become smaller than predicted by the analytical equations (for example, for $T \to 0$ and $\mathcal{E} \to 0$, with only the $|000\rangle$ state present, the variance would even go to zero).[31]

## 2.2 Linear molecules

Linear molecules differ from symmetric tops in that they display a quadratic, rather than linear, Stark effect (tops are gyroscopic, whereas linear rotors are not). The orientation cosine for the state $|J,M\rangle$ of a linear rotor is given by

$$P_1 = \omega f(J,M) \tag{18}$$

where[32,33]

$$f(J,M) = \frac{\frac{3M^2}{J(J+1)} - 1}{(2J+3)(2J-1)} \text{ for } J \neq 0, \quad \frac{1}{3} \text{ for } J = 0 \tag{19}$$

Averaging Eq. (19) over all *M* values, one finds

$$\overline{P_1} = \frac{\omega}{3Z} \tag{20}$$

(only the ground state gives a non-zero average orientation) and

$$\overline{P_1^2} = \frac{1}{Z} \omega^2 \left[ \frac{1}{9} + \sum_{J=1}^{\infty} e^{-J(J+1)/Y} g(J) \right], \tag{21}$$

where

$$g(J) = \sum_{M=-J}^{J} f^2(J,M) = \frac{2J+1}{5J(J+1)(2J-1)(2J+3)}. \tag{22}$$

The partition function is given by "Mulholland's formula" (see, e.g., Ref. 34)

$$Z = Y + \tfrac{1}{3} + \ldots \tag{23}$$

The sum in brackets in Eq. (21) can be evaluated with the help of the Euler-Maclaurin formula or, simpler, by setting the exponentials to 1 (since we are interested in the limit of $Y \gg 1$) and adding the first few $g(J)$. To leading order in $1/Y \ll 1$ and $\omega \ll 1$, one obtains

$$\text{var}(P_1) \approx 0.17 \frac{\omega^2}{Y}. \tag{24}$$

This expression is compared with the exact numerical calculation in Figs. 3 and 4, and is seen to be applicable within its range of validity. For increasingly strong fields the variance initially increases steeply to a maximum and then decreases, reflecting the onset of nearly full orientation of the dipole. This is shown in Fig. 5, where one can note both the difference in the low-field behavior and the similarity in the high-field behavior between the linear and symmetric-top rotors.

# 3. Orientation distribution function

As the previous sections make clear, one does not need to know the exact form of the probability distribution $\mathcal{P}(\mu_z)$, or $\mathcal{P}(P_1)$, in order to calculate the variance of the dipole orientation over the statistical beam ensemble. Nevertheless it is instructive to briefly discuss the distribution of orientations, because it provides intuition about the expected shape of the broadened beam and its dependence on the particle characteristics and beam parameters. As mentioned at the beginning of Sec. 2, the measured deflected beam profile is the convolution of the original beam profile (without an external field) and the orientation distribution functions, see, e.g., the detailed discussion in Ref. 17.

**3.1 Symmetric tops**

For a spherical rigid rotor with rotational constant $B$ small with respect to $T_{\rm rot}$ but large with respect to the field-dipole interaction energy, the dipole orientation distribution in the low-field limit has been shown to have the form[24]

$$\mathcal{P}(P_1) \approx \frac{1}{2}\ln\frac{1}{|P_1|}, \quad (25)$$

with $-1 \leq P_1 \leq 1$. (Alternative derivations of this equation can be found in Refs. 35,36.) The variance result (17) can be obtained directly from this $\mathcal{P}(P_1)$. The distribution (25) was derived by treating the angular momentum as a continuous variable, which is also responsible for the spurious singularity at the origin. For the general case of a symmetric top, an analogous analytical approximation for the distribution function has not been found.

To illustrate the difference from spherical symmetry, Fig. 6(a) superimposes the results of an exact quantum mechanical calculation for a prolate and a spherical top. Since for the chosen set of parameters $\mu_0\mathcal{E}$ is approaching $k_B T_{\rm rot}$, the plots are noticeably asymmetric (cf. Refs. [37,38]; for a weaker field the spherical top's distribution would approach Eq. (25)). Note that the distribution in the prolate case (bar plot) is much sharper than in the spherical case (solid curve envelope). Conversely, for an oblate top, the distribution would become broader. This variation with shape, which is found for all field strengths, is in full accord with the results of Sec. 2.1 and Fig. 2.

The histograms in Figs. 6(a-c) show how the prolate top's orientation function develops as its rotational constant increases. Two observations are of interest. The first is that for the lighter rotors [Fig. 6(c)] individual quantum states remain clearly distinguishable in the deflection profile. This is the case for asymmetric rotors as well, as demonstrated for example for the water molecule in Ref. [12].

The second observation is that all three distribution histograms are well matched by the same envelope (dashed line). Hence not only the first two moments (centroid and variance, as already mentioned in Sec. 2.1) of the orientation distribution are rather insensitive to the magnitude of $B$, but so is the entire distribution's contour (provided that $B$ remains much smaller than $k_B T$, i.e., $Y \gg 1$). This affirms the suggestion in Sec. 2.1: when faced with a need to diagonalize a large number of big matrices (this happens when both $\omega > 1$ and the temperature is high compared to the rotational constant) one may follow the strategy of artificially increasing the value of $B$. This can decrease matrix size without affecting the calculated deflection parameters, as we just saw. This offers an alternative to appealing to a classical calculation.[25,37-41]

## 3.2 Linear rotators

For linear molecules an analytical expression for the orientation distribution function can be obtained in the classical limit.

For $J \gg 1$ we can use the classical form $M = \sqrt{J(J+1)} \cos\varphi$ (where $\varphi$ is the angle between the incoming molecule's rotation axis and the field) in Eq. (19) to write Eq. (18) as[32]

$$P_1 = \frac{\mu_0 \mathcal{E}}{4 E_{rot}} \left(3 p^2 - 1\right). \tag{26}$$

Here $E_{rot}$ is the rotational energy and $\cos\varphi$ is abbreviated by $p$. This expression can also be obtained directly from a perturbative analysis of the classical motion of a linear dipole.[4]

The orientation of the molecular rotation axis is *a priori* arbitrary, hence its probability distribution is uniform: $\mathcal{P}(p)dp = 1 \cdot dp$. Using $\mathcal{P}(P_1)dP_1 = \mathcal{P}(p)dp$ together with Eq. (26) and defining $\Omega \equiv \mu_0 \mathcal{E}/4 E_{rot}$, we obtain

$$\mathcal{P}(P_1) = \frac{1}{2\sqrt{3\Omega}\sqrt{P_1 + \Omega}} \tag{27}$$

with $-\Omega < P_1 \leq 2\Omega$. Arising from a perturbation treatment of $\mathcal{E}$, this result assumes $\Omega \ll 1$. The function is plotted in Fig. 7. There is a singularity for $P_1 \to -\Omega$; an analogous rainbow singularity for polarizable linear molecules in static or oscillating fields has been noted previously.[42,43] As in the case of symmetric tops, it is a consequence of assuming a continuous distribution of angular momenta.

There is a striking contrast between the shapes of the distribution $\mathcal{P}(P_1)$ in the cases of symmetric-top (Fig. 6) and linear (Fig. 7) rotors, and between the ranges of the orientation cosine ($-1 \leq P_1 \leq 1$ for the former, and $-\mu_0 \mathcal{E}/4 E_{rot} < P_1 \leq \mu_0 \mathcal{E}/2 E_{rot}$ for the latter). This reflects once again the difference between the gyroscopic and non-gyroscopic rotational nature of the two types.

The above perturbative distribution is classical, but it should be kept in mind that within perturbation theory all of the orientation $\bar{P}_1$ of the linear dipole,[18] and much of the variance $\text{var}(P_1)$ (Sec. 2.2) actually come from the $J=0$ state. As mentioned in Ref. 18, this is related to the fact, pointed out by Pauli, that classically only nonrotating linear molecules "which execute small vibrations about a position of rest"[44,45] contribute to the polarization.

# 4. Conclusions

The broadening of a beam of polar molecules or clusters in an electric or magnetic field gradient can be characterized by the variance of the orientation (direction cosine) distribution. To assist with the interpretation of beam deflection measurements, analytical expressions for broadening in the weak-field regime (Eqs. (15), (16) for rigid symmetric-top rotors and Eq. (24) for linear molecules) have been compared with full diagonalization of the rotational Hamiltonian matrix.

Because of their distinct rotational properties, symmetric-top and linear dipoles exhibit very different low-field behavior (constant variance in the former case and vanishing variance in the latter case). However, both display a subsequent rise followed by a decrease in the strong-field limit (Figs. 1 and 5), associated with increasingly forced orientation along the field direction. This behavior is consistent with their orientation probability distribution functions, discussed in Sec. 3.

A convenient observation is that even for strong fields or small rotational constants, the dependence of the beam broadening parameter on the particle shape is found to resemble that from the analytical perturbation-theory formula (Fig. 2).

The comparisons of analytical and numerical results presented here are aimed to be of utility when cluster or molecular dipole moments and shapes need to be estimated in relation to experimental measurements of beam broadening upon passage through an electric or magnetic field.

It is also noted that within a wide range of values of the rotational constant $B$, the deflection parameters are only weakly sensitive to the magnitude of this quantity. Thus resource-intensive computations of deflection patterns may be shortened without significant loss of accuracy by working with a larger $B$ and correspondingly smaller matrices.

An important aspect of beam deflection and broadening behavior in experiments of this type is that the particles are not in thermal equilibrium within the external field, but transit through it adiabatically. The ensemble is assumed to be characterized by an initial rotational temperature acquired prior to entry into the deflecting field. It would be interesting to extend the semiclassical and quantum analysis of broadening to the case when the orientation distribution function of the rotors is not statistically distributed, as in Sec. 3, but prealigned by a laser pulse.[46,47]

**Acknowledgements**

We would like to thank Prof. Rolf Schäfer, Dr. Anthony Liang, and Dr. John Bowlan for useful discussions. The work of V.K. was supported by the U.S. National Science Foundation (PHY-1068292).

**Figure captions**

**Fig. 1.** Variance of the orientation distribution for spherical ($A/C$=1) and prolate ($A/C$=2) symmetric tops as a function of external field strength. Here and in subsequent figures, conversion to the dimensionless parameter $\omega$ can be performed via Eq. (8). For concreteness, this and other figures involve electric fields (taking $\mu_0$=1 D), but the results are equally applicable to magnetic deflections. While a spherical rotator is an artificial case for an electric dipole, it is employed here because it is realistic for a magnetic dipole and is furthermore a convenient approximation for nearly-spherical polar clusters, see text for discussion. Empty symbols show the results of calculations for the adiabatic-entry ensemble. They illustrate that for fields that are not too strong, var($P_1$) is independent of the rotational constant, field and temperature, and is determined only by the shape, as per Eqs. (15), (16) which are plotted with dashed horizontal lines. The thin solid line is a guide to the eye. Crosses denote in-field-equilibrium ensemble variances for the prolate top; significant differences are observed only at very high field strengths.

**Fig. 2.** Variance as a function of the $C/A$ (oblate) and $A/C$ (prolate) ratio for various values of the rotational constant $B$ (panels a,b) or the electric field $\mathcal{E}$ (panel c). Lines: exact calculation. Circles: first-order perturbation theory result, Eqs. (15),(16). (a) $\mathcal{E}$=50 kV/cm, $\mu_0$=1 D, $T_{\text{rot}}$=25 K. Crosses: first- plus second-order perturbation theory terms. Values of $B$=0.1 and 1.0 cm$^{-1}$ correspond to $\omega$=8.5 and 0.85, respectively. (b) $\mathcal{E}$=100 kV/cm, $\mu_0$=1 D, $T_{\text{rot}}$=5 K (a line for $B$=0.01 cm$^{-1}$ would be indistinguishable from that for $B$=0.002 cm$^{-1}$). Values of $B$=0.002, 0.1 and 1.0 cm$^{-1}$ correspond to $\omega$=850, 17, and 1.7, respectively. (c) $B$=0.1 cm$^{-1}$, $\mu_0$=1 D, $T_{\text{rot}}$=25 K.

**Fig. 3.** Variance of the orientation distribution for a linear polar molecule with $\mu_0$=1 D, $\mathcal{E}$=100 kV/cm, and $T_{\text{rot}}$=100 K. The $B$ scale is linear in the main plot and logarithmic in the insert. For this field, the condition $\omega$<1 and Eq. (24) are applicable for rotational constants $B \gtrsim 2$ cm$^{-1}$. Such a parameter range is appropriate for diatomic hydrides; for weaker field and/or smaller dipole moments, Eq. (24) also can describe other molecules with lower rotational constants.

**Fig. 4.** Solid line: the exactly computed variance of the orientation distribution for a linear polar molecule with $\mu_0$=1 D, $B$=1 cm$^{-1}$, and $T_{\text{rot}}$=50 K. For low electric fields it follows that of Eq. (24) (dotted line), but then it saturates and bends over, as shown in Fig. 5.

**Fig. 5.** Variance for a linear molecule, with $B$=1 cm$^{-1}$, $\mu_0$=1 D, and $T_{rot}$=5 K. Adiabatic-entry and in-field thermal equilibrium results overlap until the fields become quite large. Note that the variance vanishes for zero field, in contrast to the symmetric-top behavior in Fig. 1, but at high fields both rotor types exhibit a rise towards a maximum followed by a decrease associated with transition to a field-aligned pendular state.

**Fig. 6.** Histograms of probability distributions of the dipole orientation, $P_1=\mu_z/\mu_0$, in a field of 100 kV/cm for prolate symmetric top molecules with $\mu_0$=1 D and $T_{rot}$=5K and different rotational constants $B$. In the diagram in (c) one can distinguish the contributions from individual quantum rotational Stark states. The plots reveal that the average, the variance, and the contour (dashed line) of the distribution are all practically insensitive to $B$ (provided that $B<<k_B T_{rot}$). Panel (a) also shows that the distribution for a spherical top (solid line, which is the envelope of the corresponding histogram) is wider than that of the prolate one, in agreement with the trends illustrated in Fig. 2.

**Fig. 7.** Orientation probability distribution, Eq. (27), for a classical linear dipole in a weak external field. The relevant scale for the direction cosine in this situation is set by $\Omega$, the ratio of the dipole's energy in the electric field to the rotational kinetic energy of the linear molecule. As discussed in the text, this differs from the $-1 \leq P_1 \leq 1$ range for symmetric tops.

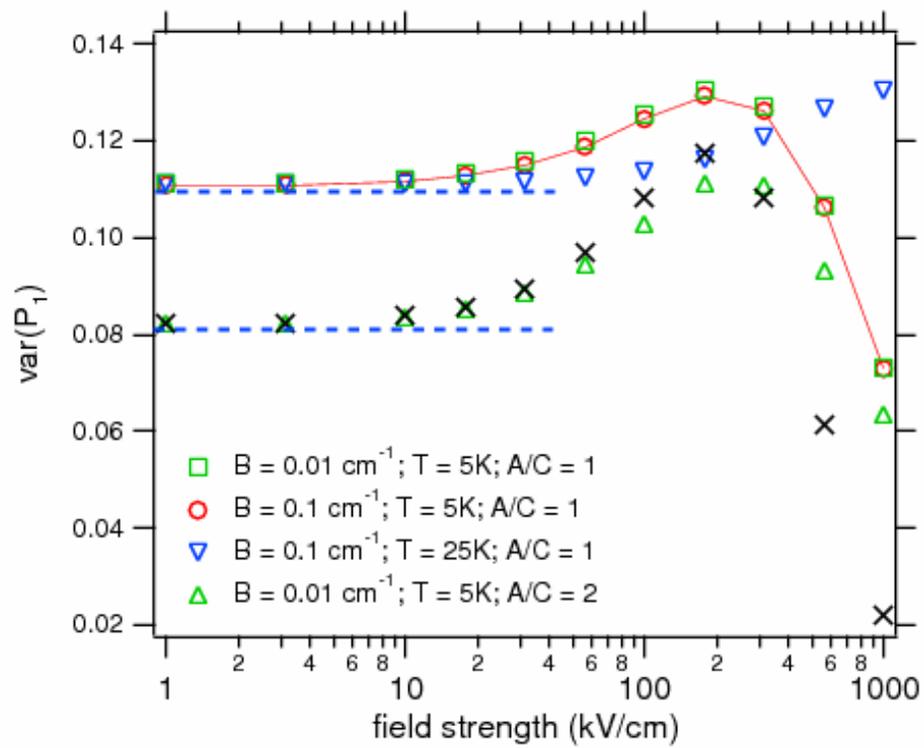

**Figure 1**

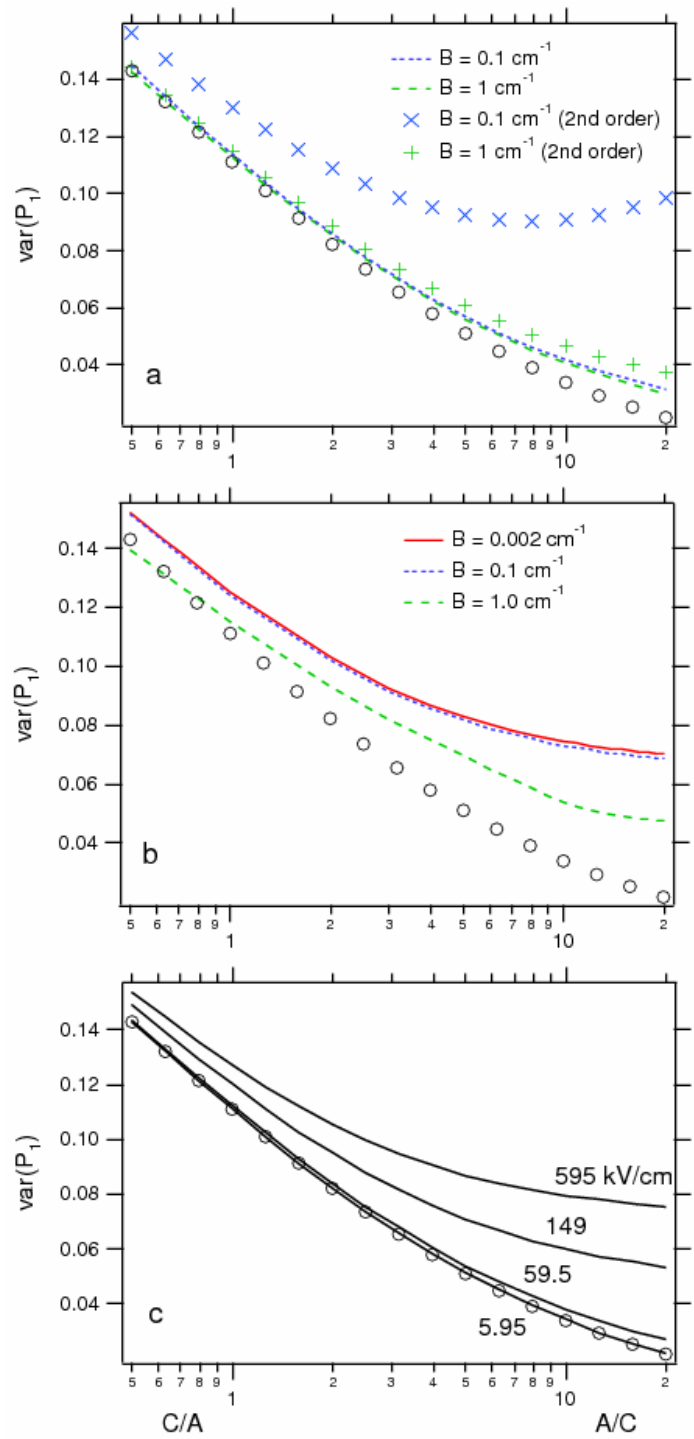

**Figure 2**

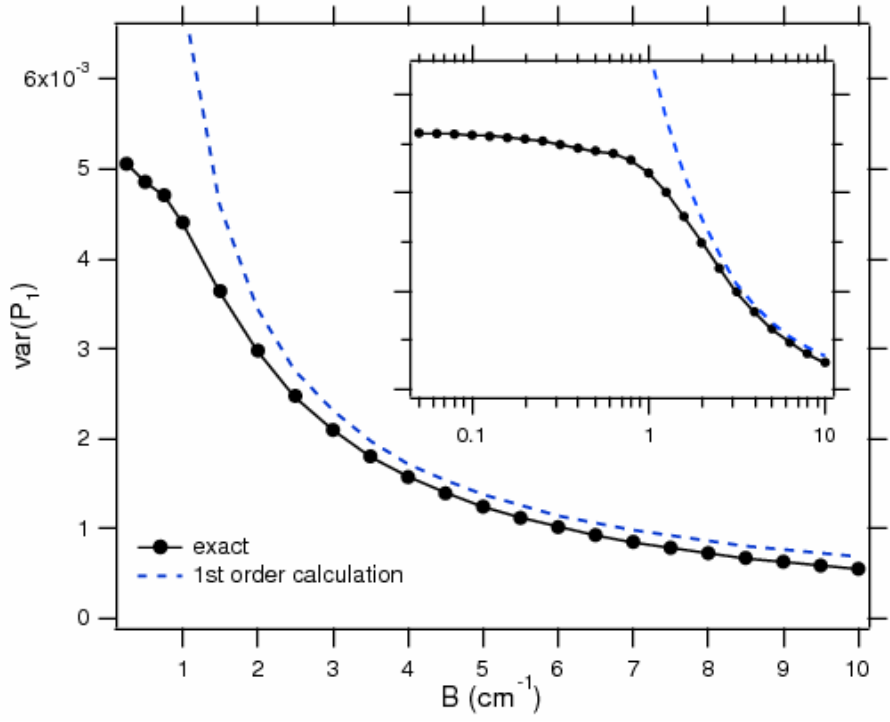

**Figure 3**

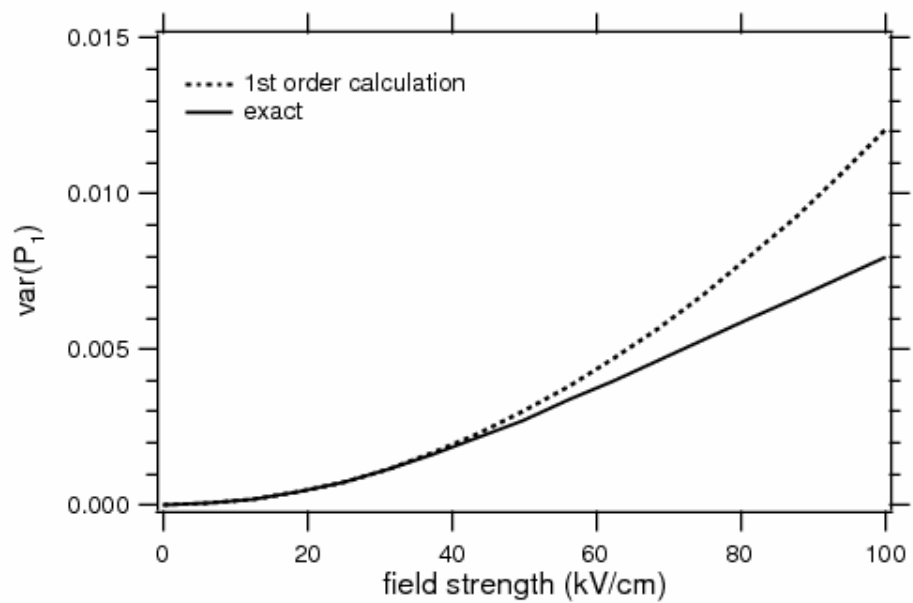

**Figure 4**

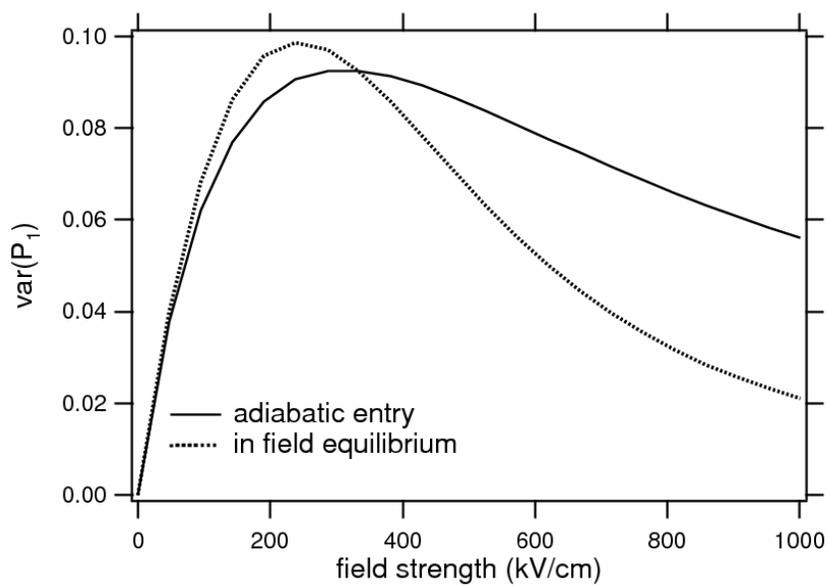

**Figure 5**

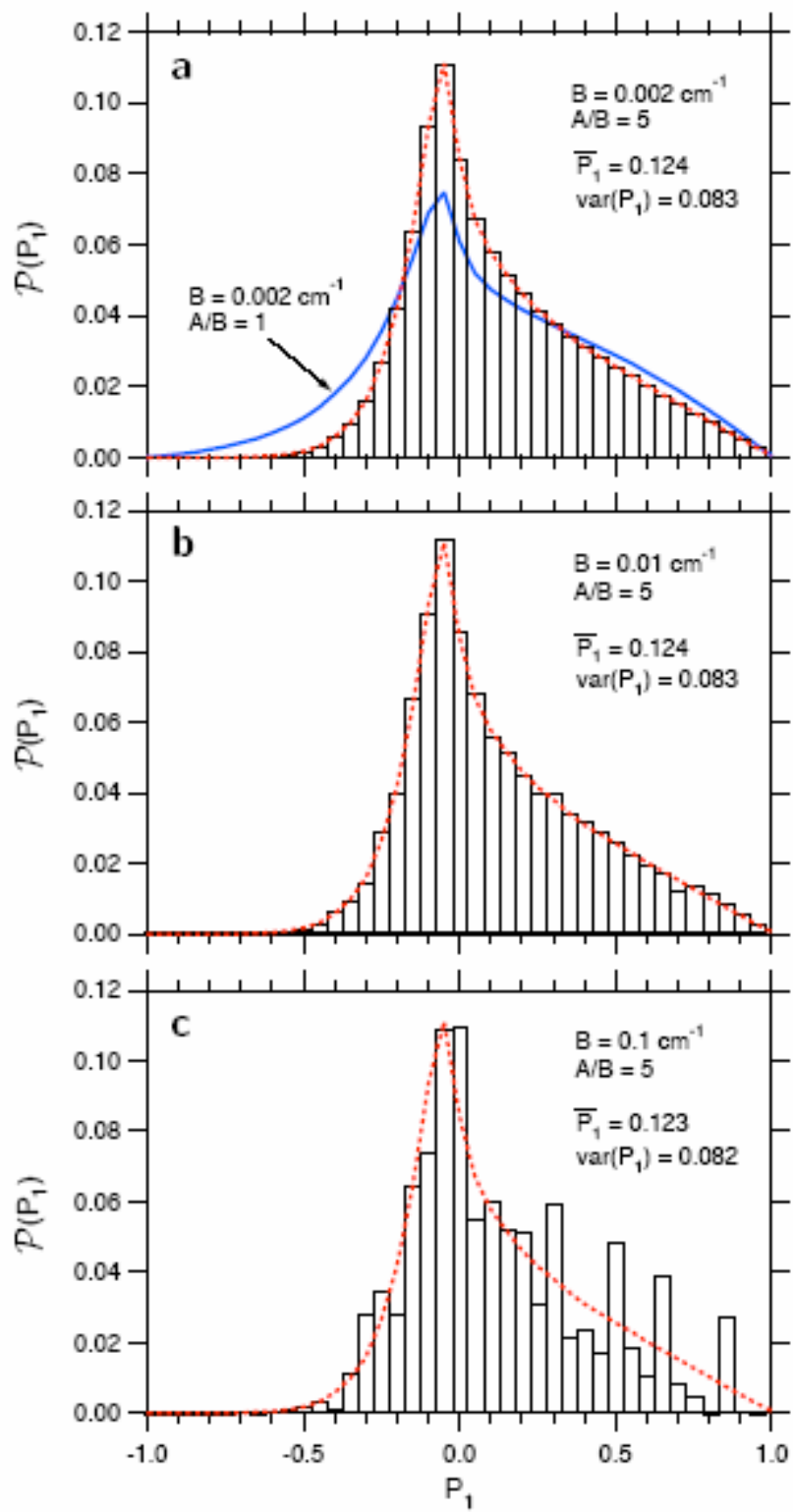

**Figure 6**

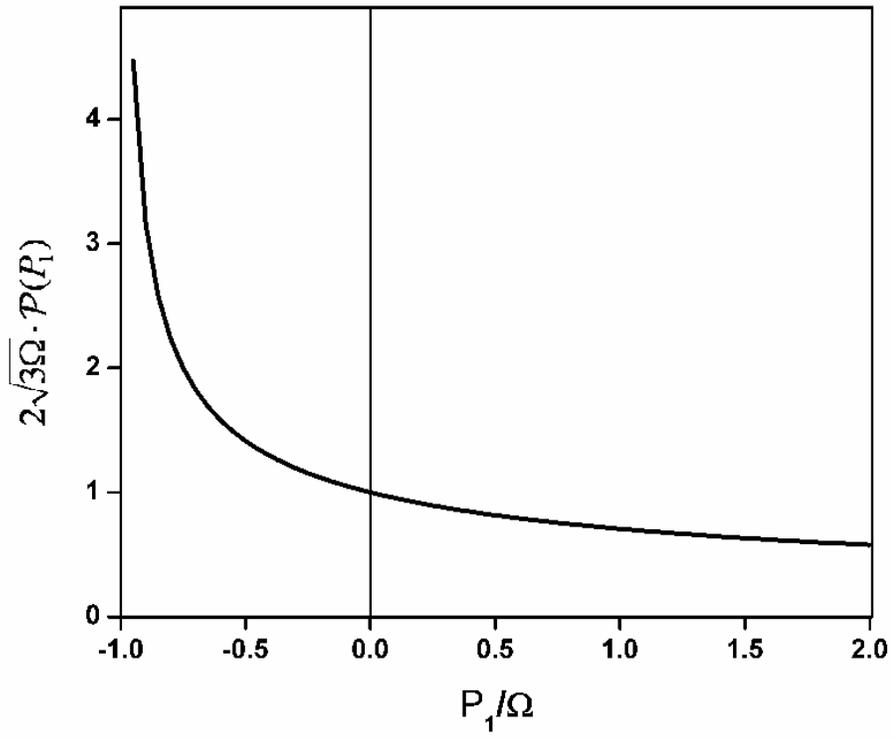

**Figure 7**